# The EPN-TAP protocol for the Planetary Science Virtual Observatory

S. Erard [1], B. Cecconi [1], P. Le Sidaner [2], J. Berthier [3], F. Henry [1], M. Molinaro [4], M. Giardino [5], N. Bourrel [6], N. André [6], M. Gangloff [6], C. Jacquey [6], F. Topf [7]

[1] *LESIA, Observatoire de Paris/CNRS/UPMC/Univ. Paris-Diderot*
5 pl. J. Janssen 92195 Meudon, France. email: stephane.erard@obspm.fr

[2] *DIO-VO, UMS2201 CNRS, Observatoire de Paris*
61 av. de l'Observatoire, 75014 Paris, France

[3] *IMCCE, Observatoire de Paris/CNRS*
61 av. de l'Observatoire, 75014 Paris, France

[4] *INAF - Osservatorio Astronomico di Trieste*
via G.B. Tiepolo 11, 34143 Trieste, Italy

[5] *INAF - Istituto di Astrofisica e Planetologia Spaziali (IAPS)*
Via del Fosso del Cavaliere 100, 00133 Roma Italy

[6] *CDPP, IRAP/CNRS/Univ. Paul Sabatier*
9 avenue du colonel Roche, 31068 Toulouse, France

[7] *Space Research Institute, Austrian Academy of Sciences*
Schmiedlstrasse 6, A - 8042 Graz, Austria

**Corresponding author:** S. Erard, LESIA, Observatoire de Paris
5 pl. J. Janssen 92195 Meudon, France.
email: stephane.erard@obspm.fr
Tel : (33) 1 45 07 78 19

## Abstract

A Data Access Protocol has been set up to search and retrieve Planetary Science data in general. This protocol will allow the user to select a subset of data from an archive in a standard way, based on the IVOA Table Access Protocol (TAP). The TAP mechanism is completed by an underlying Data Model and reference dictionaries. This paper describes the principle of the EPN-TAP protocol and interfaces, underlines the choices that have been made, and discusses possible evolutions.

# 1 - Introduction

EPN-TAP is a VO data access protocol designed to support Planetary Science data in the broadest sense. It is intended to access data services of various content, including space-borne, ground-based, experimental (laboratory), and simulated data. It is designed to describe many fields, from surface imaging to spectroscopy, atmospheric structure, electro-magnetic fields, and particle measurements. EPN-TAP is an essential part of the Planetary Science Virtual Observatory (VO), because no prexisting protocol was able to access such a large realm of data (see Erard et al. this issue, companion paper).

The EPN-TAP protocol is directly derived from IVOA's Table Access Protocol (TAP) [6], a protocol to access data organized in tables, here adapted to Planetary Science. EPN-TAP is an extension of TAP with extra characterization derived from a Data Model — similarly to ObsTAP, which is an extension based on the ObsCore Data Model [7].

The Europlanet Data Model was defined to describe many types of Planetary Science data with a standard terminology [5]. EPN-TAP uses a subset of this terminology to define standard query parameters. This subset of the Europlanet Data Model is called EPNCore. Some of these parameters are adapted from the PDAP protocol of IPDA [1] and from the SPASE protocol (Space Physics Archive Search and Extract).

Since EPN-TAP is TAP compliant, the discovery of all EPN-TAP data services can be performed using an IVOA registry. EPN-TAP services are described accurately by IVOA registries that include the TAPRegExt extension (see companion paper). Once declared in a registry, EPN-TAP compliant data services are most efficiently queried with a specific EPN-TAP client such as the VESPA tool at VO-Paris.

This paper provides a synthetic description of EPN-TAP and discusses the choice made during its definition. EPN-TAP definition includes:
- A general framework to implement data services (SQL database, the presence of the epn_core view, etc).
- A set of parameters describing the resources and their content (the EPNCore DM), plus optional parameters and attributes.
- A convention to provide numeric parameters in standard form (units/scales, etc) for the query mechanism.
- A set of reference sources to encode the string parameters (e. g. target names, etc).
- A set of UCDs defining the parameters in use in the VO context.

# 2 - Main concepts of EPN-TAP

EPN-TAP is an extension of IVOA TAP and is compliant with the TAP standard. It typically uses the TAP mechanism [6] with synchronous or asynchronous queries; VOSI for capability and metadata access [3].

TAP is a protocol dedicated to access relational database tables. It uses ADQL (the Astronomical Data Query Language, [10]) to query the databases. To allow similar queries on all EPN-TAP services, we will assume that the EPNCore data model is implemented in the database as a view (i. e., a table presenting the parameters). In order to be accessed through EPN-TAP, all databases must therefore include a view called epn_core, which contains at least all the parameters described in section 3.1. The epn_core view mainly contains a list of the "granules" available in the database, typically an entry/line for each data element, and is used as a catalog of the accessible content. The parameters describing the granules are mostly related to data description and to the main axes of variation.

## 2.1 Axes description

In practice, the user writes his query on a client interface. The client sends a formatted query to the server. The server in turn looks for matches in the epn_core view and sends back an answer. This process is illustrated in Fig. 1. A standard situation is to search data located in space, time or wavelength, therefore to issue a query based on axis coordinates.

In order to handle the multiplicity of situations, most parameters are normalized in the protocol, regardless of the content of the databases. For instance, a spectroscopy database may provide measurements on a wavelength scale in microns, while the user wants to query the data on a wavenumber scale in cm$^{-1}$ (Fig. 1). A common description must therefore be used, which should not interfere with the way the data are described, nor with the way the user wants to query the data.

The EPN-TAP query standard defines the scales and units used for parameters — e. g., the spectral axes are always described on a frequency scale in Hz. Since the databases do not necessarily use the standard scales/units internally, the epn_core view also has the function to provide the parameters in the expected units once for all. This avoids on-the-fly conversions on the database side, while the data themselves may remain in native form (Fig. 1). This view is used as an interface for the client, and can remain hidden to the user.

Similarly, the client interface may propose a variety of scales/units to the user, and convert them in Hz to write the query. It is therefore essential that such transforms are exactly reciprocal on both sides of the query system. A similar system is used for many parameters, e. g., time scales are provided in Julian Days.

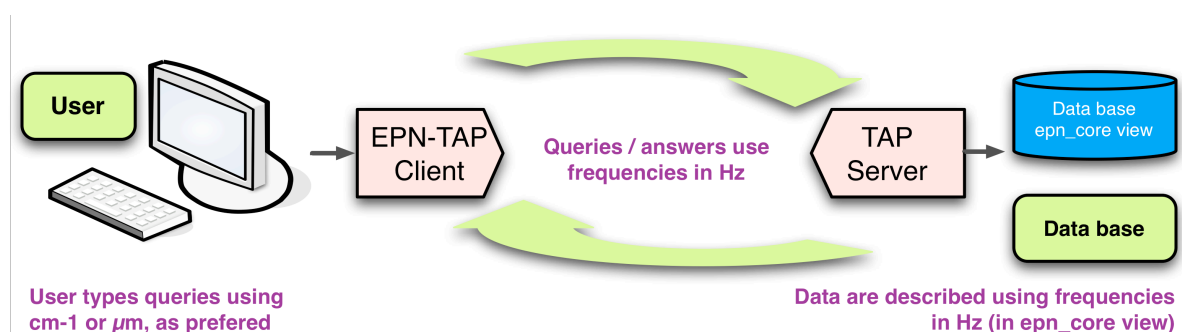

Fig. 1: Practical implementation for EPN-TAP queries. On the service side, only the epn_core view is converted to standard scales and units.

The EPN-TAP protocol is closely related to the TAP protocol, and mainly differs by the definition of its core parameters. The server side relies on a general framework for TAP, while the client performs most EPN-specific operations and turns them into fully TAP-compliant queries, which can be handled directly by the service framework through ADQL.

Parameter names are mostly used as tags to pass the values between the client and the server. Since they are used to handle a variety of situations, science fields, etc, they may not reflect the exact meaning of the parameters in the frame of a specific database. This again is not an issue, since parameter names are not normally seen by the users (depending on the client interface).

A particular situation arises with the spatial coordinates, because of the extreme diversity encountered in Planetary Science. In order to simply formulate a query, the general type of coordinate system (e.g. celestial coordinates, geographical coordinates, Cartesian coordinates in a volume, etc) must be known in advance. For this reason the description (provided by the spatial_frame_type parameter) must be included in the column description of the TAP response [6] and in the metadata returned by the service. In the future, this will used to select services in advance. However, only the parameters of the epn_core view can be used for data selection in a TAP query; therefore important service attributes are best stored as parameters even when they remain constant throughout the table (e. g., the same spatial_frame_type parameter can be used to select granules individually).

## 2.2 Data description

Apart from the data description, the epn_core view may include the data itself, or links to data files. The data structure is not necessarily constant among all granules in the epn_core view, and a service can contain a mix of images, spectra, etc.

In addition to the granules defined above, at least one "dataset" entry is required for each service. Parameters describing "datasets" provide the range encompassed by their elements/granules, e.g. coordinates or observing dates. "Datasets" and "granules" entries are identified using the resource_type parameter. A query on "dataset" may be used to return only global information on a service, without a long list of available data products, and is therefore the preferred access mode in discovery phase. For this reason, an EPN-TAP client will preferably default to resource_type = dataset. In the epn_core view, datasets are best located at the beginning for visibility: most VO clients only load a limited number of entries by default, so the last ones are often not displayed.

Additional "datasets" can be defined inside the epn_core view. Such datasets consist in subsets of granules selected according to various criteria by the data provider. A complex PDS data set for example can be sliced into several subsets accessed independently through EPN-TAP, e. g., to identify different processing levels. This allows data providers to make their data available in EPN-TAP without going through the burden of generating alternative versions of their databases.

An important part of the service design is related to the identification of the granules, and is left to the data provider. The simplest situation corresponds to one entry per data file, but complex situations may call for other solutions. For instance, if an image contains both Mars and Phobos, the basic approach is to have one granule with the two target names stored in the target_name parameter. Alternatively, if the target is considered as the main entry, there could be two granules (Mars and Phobos) pointing to the same image file; this permits to provide the coordinates relative to each body with no ambiguity (a similar situation may occur when the data files contain several data products of different types). A third possibility would be to combine the first two, and to define three granules pointing to the same image. Although there is no mandatory rule, this third possibility is in general not desirable: redundancy in the epn_core view will result in duplicate answers, which may be both confusing and unpractical for the user. Data providers will in general want to give answers as explicit as possible.

## 2.3 Writing and matching queries

Altogether, the epn_core view is composed of many fields (Fig. 2): all mandatory EPN-TAP parameters; possibly optional or extra parameters; data access information, either data embedded in the view or access information to data files.

In the most general case, queries are written from a client and sent to all accessible services. Queries must therefore respect the standard: only mandatory parameters can be queried, and are used as filters. Services receiving unknown parameters would respond with an error code. Conversely, parameters not present in the query are not used to filter the response.

When receiving a query, the server looks for matching lines in the epn_core view (Fig. 2). The answer is an excerpt of the view containing all its columns, including the EPNCore parameters and possibly the data, embedded in a VOTable. Data access is therefore provided according to the table definition.

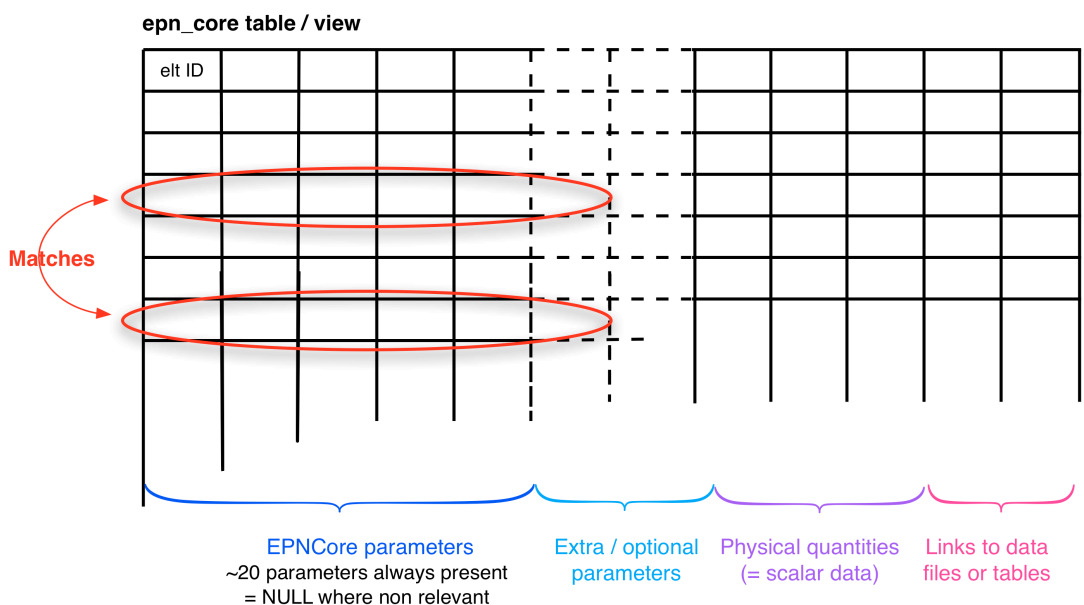

Fig. 2: Query of the epn_core view and returned values

When only one service is addressed, the VOSI mechanism provides access to the list of fields in the epn_core view. Once this is known, any table field can be queried with TAP, including optional and non-standard parameters, plus the data themselves when they are contained in the epn_core view. This mechanism provides a complete access to the data service content (in contrast to the PDAP protocol v1, for example).

## 2.4 ObsTAP versus EPN-TAP

Close inspection would confirm deep similarities between the two protocols. Since EPN-TAP parameters accept more values than ObsTAP, it could be interpreted as an enlarged version of ObsTAP. However this is not the way it is intended, and we stress the need to implement ObsCore on simple but essential data services, which are also valuable for Planetary Science.

Examples of such applications include the following use cases:
- The user needs to retrieve a list of the brightest celestial IR sources in the whole sky to check for stellar occultation from a spacecraft. Practically, he has to identify a catalogue in Vizier including the adequate quantity, say K magnitude. The catalogue may be sorted in Vizier web interface (not always possible) or transferred to TOPCAT for visualization and analysis (although there may be difficulties e.g. with coordinate format).
- The user needs to get scaled IR spectra of reference stars. A current solution is to get a list of reference stars, to retrieve their spectral type and magnitude in a given band, to grab spectra of similar spectral types either at ESO or IRTF, and to scale them to the magnitude of the targets.

Although simple, such use cases may be remarkably difficult to implement for the casual user. One of the hard points is the difficulty to identify services distributing the required physical quantities, the reason why ObsTAP was initially setup. The target_class and o_ucd parameters in ObsTAP help solve this problem, as their counterparts would in EPN-TAP.

ObsCore is obviously very efficient to distribute simple services with mainly one quantity documented for many targets, and its use for astronomical services in support of Solar System observations is encouraged. We cannot stress enough the need to implement use cases or services in Astronomy to support the observation of the Solar System.

## 3 - EPNCore / EPN-TAP parameters

The TAP mechanism is used here with a set of specific parameters. The mandatory EPN-TAP parameters constituting the EPNCore have been defined on the basis of real use cases in various fields, so as to handle most data services related to Planetary Science (see list of first services in companion paper).

EPN-TAP can also query parameters not included in the EPNCore. Some of these parameters are defined precisely but are relevant only to very specific data services. Those are not mandatory, but they must be implemented as defined in the standard when present. Beside, the names of optional parameters are reserved for this particular usage and must not be used to introduce other quantities.

In addition, several optional attributes can be used to define general properties of the service itself, such as a detailed description of coordinate systems in use, the processing level of the data, or a description of the service. Such attributes can also be used as parameters describing all the granules of a service so they can be grabbed by TAP, but their values are in general expected to remain constant.

Although EPN-TAP bears many similarities with the ObsTAP protocol, large variations are used to handle the specificities of Planetary Science data. In some instances the ObsCore parameter names have been preserved, but in general the acceptable values are different or constrained differently, and the meaning of the parameters is therefore slightly different. Their names have then been changed to avoid any confusion, since in principle both EPN-TAP and ObsTAP services can be queried with the same client. Other concepts have been adapted from the PDS and from SPASE.

## 3.1 Parameters

EPN-TAP parameters can be grouped in several categories: axis ranges, data description, and data access.
- Axis range parameters provide the data coordinates in space, time, spectral domain, and photometric domain. They allow the user to focus on particular ranges along these axes.
- Data description parameters document the data in a more general way, providing target description (name and type), data origin (instrument and facility plus references), and basic description of the data themselves (data and measurement types). The latter two parameters allow the user to find particular types of data, e. g. surface images, vertical atmospheric profiles, or spectral cubes.
- Data access parameters provide links to the data files, or in some cases the data themselves. All those are optional parameters.

The complete list of parameters is given in Table 1. All EPN-TAP parameters are documented with a numerical type, unit, UCD, and description (free field), according to VOSI specifications. This information is available in the service response and can be used by the client. Expected values are listed in Table 1.

### Axis parameters

Spatial axes are of course more intricate than in Astronomy, given the wide variety of coordinate systems in use. A particularity of EPN-TAP is to use a spatial_frame_type parameter that provides the "flavor" of the coordinate system and defines the nature of the spatial coordinates (Table 2): either celestial (right ascension and declination + possibly distance), body-fixed (longitude and latitude + possibly elevation), Cartesian (distances), cylindrical, spherical, or healpix for more general situations.

The 3 spatial coordinates are defined according to the previous parameters, i.e. their meaning and physical dimension are context-dependent. In addition to the coordinates, the spatial resolution is provided. The exact coordinate system in use is documented through optional parameters spatial_coordinate_description and spatial_origin.
Although the meaning of the latter is rather straightforward, the spatial_coordinate_description is more tricky: it is expected to provide complete reference to the system in use and its properties, including: target body; reference ellipsoid or shape model; control point network; latitude definition (planetocentric vs planetographic); orientation (east- vs west-handed). In practice, the use of a comprehensive list of possible systems in preferable. A simple acronym such as Mars_IAU2000 would then define the coordinate frame completely. Although the IVOA STC [13] also aims at providing standard references to coordinate frames, and actually includes some frames in the Solar System, it was not found flexible enough for the need of our community. A specific reference list of frames is therefore being compiled by the Europlanet group from existing international standards. When ready, it will be submitted to IAU for approval.

Other data axes are handled more simply. Time is accompanied by a time_sampling_step and an exposure_time parameters; the latter provides the time resolution of observations while the former is used to document regular time series found e.g. in plasma measurements. An optional time_orign parameter can used to specify the place where time is measured, to account for light-path differences when comparing event-based measurements (the definition of "UTC" does not cover this). This is in particular required when comparing ground-based and space borne data.

Parameter spectral_range is accompanied by a spectral_sampling_step and a spectral_resolution parameters, the latter providing the Full Width at Half Maximum of the measurements. Documenting both resolution and sampling step of the axis parameters is important to handle generic queries in different fields. Similar optional keywords are also reserved to provide spectral values for particles (e. g. mass spectroscopy, which uses different units).

Photometric axes are defined in a similar way by documenting the range of the main three angles (incidence, emergence and phase). These parameters will allow the user to search data related to surface reflectance and emissivity, or radiative transfer in the atmospheres, including simulation and laboratory measurements.

All axis parameters exist in two versions to provide minimum and maximum values, so as to define a range for searches. Whenever only one value is relevant, it should fill both min/max parameters.

## Data description parameters

Targets are referred to by name and class. Possible target classes are limited to a finite number of values, the list of which is part of the standard (Table 3). Target classes not only allow the user to search for a class of object, but also remove any ambiguity between homonyms (e.g. Io can be either a satellite or an asteroid). In contrast with ObsCore, the target name is a crucial search parameter for EPN-TAP, since it is the only possible way to identify a Solar System object. Its use is enlarged to samples when unambiguous (e.g. for meteorites, lunar samples, and other samples for space missions such as Stardust).

The dataproduct_type parameter is similar to that of ObsCore, and provides the high-level science organization of the data (Table 4). Although larger than ObsCore's, the list of possible values is again limited and is a part of the standard.

The measurement_type parameter is more similar to the o_ucd parameter of ObsCore, since it also provides the UCD of the main physical quantity in the data service. This is not always defined for Planetary Science (see below).

Two other descriptive parameters provide references to the instrument generating the data: one for the instrument itself, the other for the "instrument host", i. e., either a spacecraft, a telescope, or a ground-based facility.

The resource_type parameter distinguishes between granules and datasets, i.e. sets of granules. Description parameters for datasets provide the complete set of values for their granules. When several datasets are present, the dataset_id parameter will permit to restrain queries to identified datasets, and will provide cross-reference between datasets and individual granules.

The index parameter provides a unique line number in the epn_core view. It is introduced as an EPN-TAP parameter so as to permit cross-references in the database after a first query. This solution is preferred over an internal database index, which may not remain constant when updating the content.

### Data access parameters

A set of optional parameters is available to provide URLs to data files. They are related to the formatted data files, to previews in standard image formats, or to the original files whenever those are distributed in unusual formats. Additional parameters are available to provide file size, format and name. Although the data formats must be described in the epn_core view, support of data formats is not part of the protocol and is left to the visualization tools. The VESPA client uses the preview URL for quick-look visualization of the data in its web interface, and may select destination tools according to data format in some cases. Finally, the file_name parameter is intended to provide reference to the granule, but also to search services that encode information in the file names themselves.

### Other parameters

Other optional parameters are available for various purposes.

Right Ascension and Declination are always available to provide celestial coordinates in addition to the main coordinate system, whereas it is redundant or used to provide other coordinates (e. g. coordinates of the target in celestial images, whereas the main coordinates are used to identify the region observed). This may be handy for instance when sending data to TOPCAT, which uses them directly in plots. A target_distance parameter is also available to document the observer's distance, be it a spacecraft or an Earth-based telescope.

The Ls parameter may be used to store the heliocentric longitude of the target, which is a standard measurement of the season, and is particularly useful to study atmospheric phenomena.

The processing_level parameter is similar to ObsCore calib_level. In EPN-TAP however, 6 calibration levels are defined to accommodate derived products, especially in imaging and mapping. They follow the CODMAC nomenclature used in space data archives (Table 5).

The element_name parameter can be used to introduce feature names on a planetary surface, while target_region is reserved for generic names. They are similar to target_name and target_class for local features, including global regions in Solar System bodies such as "atmosphere", "ionosphere", etc

The species parameter can be used to introduce simple molecular formula, such as $H_2O$ or $CO_2$, typically when providing chemical abundances in an atmosphere. Case must follow the standard chemical syntax. For more demanding purposes, InChiKeys may for instance be provided in a specific parameter, but their use is not supported in EPN-TAP.

Finally, the "reference" parameter can be used to introduce bibliographic references, typically with a Bibcode.

## 3.2 Units

The epn_core view must provide all quantities in the EPN-TAP conventional scales and units to make universal queries possible across different datasets — this is mostly relevant to axes definition. Honoring this convention does not involve any conversion in the database itself, though. On the other side, an EPN-TAP client should ideally allow the user to enter his preferred scales and units, and convert them to the EPN-TAP standard using the symmetrical conversions.

Concerning spectral quantities, the EPN-TAP convention is to provide spectral quantities as frequencies measured in Hz, assuming propagation in vacuum. Whenever convenient, the native values can be provided through specific parameters. This may prove helpful when the data are passed to specialized tools such as CASSIS. Beyond unit conventions, it may be stressed however that few VO spectral tools are currently adapted to the need of Planetary Science, which often deals with reflectance spectra.

Spatial coordinate have units related to the type of frame coordinate in use. They are usually provided in degrees/minutes/seconds or Astronomical Units. Longitude and latitude ranges and north-pole orientation follow the IAU convention (which is not intuitive for small bodies); longitudes are always increasing eastwards [12].

Times differences are provided in seconds, while dates are provided in Julian days as double precision floats to maintain acceptable accuracy (~1 ms).

## 3.3 References for string values

Non-quantitative EPN-TAP parameters cannot take arbitrary values either. There are typically two cases:
- Values have to be selected in short lists related to the standard definition. This is the case for target_class, dataproduct_type, or resource_type.
- The parameter is associated to a reference list, in which values must be selected. This is the case for target_name, which values should be retrieved from the official IAU nomenclature.

General parameters are referred to the IAU thesaurus (target_region) or IAU nomenclature [16, 17] (target_name, element_name). Target names are particularly sensitive since most objects have several names. The official name is expected in the target_name parameter, with proper case. Case is therefore currently an issue, because the ADQL standard does not support case variations. In practice however, case-sensitivity is supported in some VO frameworks (in particular in DaCHS) and the EPN-TAP standard imposes it. Honoring the ADQL constraints would otherwise force data providers to use lower cases and therefore include non-standard names in their epn-core views. As a help to query writers and data providers, the SSODnet name resolver at IMCCE [11] provides the official IAU name of many Solar System bodies with the correct case; the SSODnet database is constantly updated with new object names [16] and by integrating older archives. Currently, services containing data of interest might not be visible if they do not use the recommended IAU nomenclature for planetary bodies. In the long run, a system completely avoiding cases on both sides (client and server) is required to support all situations. Another general issue related to the current version of ADQL is the difficulty to handle multiple-valued string fields.

Other parameters are difficult to refer correctly at present, including instrument_name and instrument_host. Although several institutions provide lists of applicable values (e.g., IAU, PDS, Spice, NSSDC), these sources are not comprehensive at present. For instance, the IAU list of ground-based observatories does not include radio-telescopes, and the PDS list does not include orbital telescopes or spacecraft mostly devoted to Astronomy or Earth observation. A project for the Planetary Science VO is therefore to complement these lists, provide a conversion table between various sources (e.g., Spice and NSSDC), and submit them to IAU for approval. Using the CCSDS registry for space missions (SANA) may be an alternative.

As mentioned above, the measurement_type parameter introduces a UCD for the main quantity in the data service. EPN-TAP uses "UCD1+" from the current IVOA list [8]. However, UCDs in Planetary Science are often not defined, e. g., those related to reflected light or in-situ measurements. A possible enlargement of this list is currently discussed in the IVOA and IPDA working groups. Another difficulty encountered here is that the UCD is related to the physical quantity, not to the type of observation performed (e. g. phys.absorption;em.opt.I is eligible, while stellar_occultation is not). Therefore, some types of measures cannot be searched currently. This may be solved in the future by adopting an observation_type parameter, which is however difficult to define.

## 3.4 Data structure

An EPN-TAP service can contain four types of data: (a) scalar data fields in limited number; (b) data contained in one separated file (image, table, etc); (c) data spread on several separated files; (d) data computed on the fly. These situations may be handled as follows:

(a) The data may be included in the epn_core view as separated columns with specific, non-standard parameter names. In general dataproduct_type = ca (catalog) is appropriate, and no access_* parameter is needed in the epn_core view. Units and dimensions may be provided in the response VOTable (e. g., using the q.rd definition file of DaCHS).
(b) A URL to the external file is provided on each line through the access_url parameter, so that the client can easily download the selected files. This description may be completed by the access_format, access_estsize, and preview_url parameters. The dataproduct_type parameter must be filled according to the data organization type (e.g., image, time_series, etc).
(c) A "main data product" must be identified, which is described as in (b). Additional data products are linked and described using parameter names derived consistently from the standard ones. The parameter preview_url is actually a common example of such a situation. Other examples include images with associated ancillary data in separated files, referred to as e.g., ancillarydata_access_url, and alternative output format referred to as native_access_url.
(d) The access_url must point to a computing system that will process the query, e.g. forwarding a query to a computing service with adapted parameters.

# 4 - Setup

## 4.1 Service implementation

EPN-TAP services may be implemented in various ways. The first ones have been installed using the GAVO/DaCHS framework; some have been installed successfully on VO-Dance. In addition to the DaCHS installation document [4], tutorials to install EPN-TAP services using DaCHS are available [2].

DaCHS normally expects a PostgreSQL database, but can support MySQL or NoSQL databases through PostGreSQL's foreign data wrapper. The database does not have to be located on the same machine as the framework. This allows the data provider to quickly set up a service from an existing database, with no conversion or duplication (this has been done for the HELIO services and M4ast).

When starting from scratch however, building a PostgreSQL database is the most convenient way. Several methods have been used for the test services at VO-Paris, mostly based on IDL/GDL routines, which provide the only versatile interface with PDS3. In many cases the data files must be opened and read to retrieve information about the granules (PDS3 or FITS headers). A dataset catalogue, if complete enough, may suffice to build the database and the epn_core view. A VOTable can be also be used to build the database, e. g., through TOPCAT jdbc extension.
Again, all EPNCore mandatory parameters must be present (but can be left empty) and provided in the correct unit. At least one dataset line is required, which summarizes the whole database. IDL routines writing the database and view, together with templates containing generic definitions of the mandatory parameters, are available to help defining new services.

In the DaCHS framework, services are defined in a file "q.rd" that maps the epn_core view. It contains the list of parameters present in the view, each associated to its attributes: numerical type, unit, UCD, and a short description string. Units are defined according to IVOA [15] and IAU [12] standards.

Like every IVOA service, EPN-TAP services are identified in an xml file providing the declaration to the registry. This file contains a description of the service and its content, and references to the TAP standard [14]. It is used by the client to connect to the available services and to display some indication of their content. This information is not reachable by the TAP mechanism and is not included in the service response.

## 4.2 Clients

EPN-TAP services can currently be queried in several ways:
(a) The VO-Paris VESPA client [20] can be used to query services declared in the OV-Paris registry, or to access local services not yet registered by providing their URL. The EPNCore parameters are entered in the user's preferred unit scales, and converted to EPN-TAP standard. Selected results can be sent to IVOA visualization tools through SAMP.
(b) The TOPCAT tool may be used as a low-level client to send general TAP queries to individual databases, visualize data, and make data available to other clients through SAMP [19].
(c) The DaCHS framework includes a client (ADQL query page) which permits to send general TAP queries to local databases individually. This is mostly intended as a maintenance facility for local databases.

An EPN-TAP client may set a default value for some parameters, in particular for resource_type. A query using a single parameter resource_type = granule would reply with the complete list of granules / data files in the service, which is not optimal for resource exploration. Setting the "dataset" value alone will return a limited number of matches per service (but at least one) and is the preferred way to list the available services; it may be the client's default.

VESPA implements various query modes. The standard one is to provide a generic EPN-TAP interface to write and send queries to all EPN-TAP registered; only mandatory parameters are supported, so that all services are expected to answer correctly. The result of a general query is a page displaying the number of answers from all reachable services; currently the user has to select one service to access its specific answers.

An "Advanced query form" is also available on the service results page. For a given service, it provides the same interface completed with optional and specific parameters, which are retrieved through the VOSI mechanism. This allows the user to query a specific database using all its parameters.

Finally, the "Custom resource" mode of VESPA can connect a non-registered service given the server URL and the schema name. This allows for testing a service that is not yet registered.

# 5 - Queries and responses

## 5.1 EPN-TAP queries

A TAP query consists in looking for certain values of the parameters in the data table. Its arguments are therefore the parameters/columns of this table. Such queries use parameters as filters on the database contents, and return only the lines of the table matching the arguments.

The client must use the HTTP GET or POST protocols to send queries to services. The query is composed of the URL of the service, and ADQL language [10] is used to express the request. The TAP query is very generic and there is no mandatory parameter associated with it. A typical query is the following:

*http://<server address>/tap/sync/request=doquery & lang=adql & query=select * from epn_core where time_min > '2455197.5' and time_max < '2455927.5'*

This will return all kind of data from 2455197.5 (01/01/2010) to 2455927.5 (01/01/2012) in Julian days (target is not specified).

Some parameters can be multivalued in the sense that the epn_core view can accommodate several values, in particular when related to datasets. The separator between values is always a space. To query such parameters, the "like" operator must always be used instead of the "=" operator. These fields include: target_name, target_class, instrument_name, instrument_host_name, measurement_type.

*http://<server address>/tap/sync/request=doquery & lang=adql & query=select * from epn_core where time_min between '2455197.5' and '2455927.5' and target_class like 'comet' and target_name like '1P'*

The service will return all data of any type for comet Halley (1P) from 2455197.5 (01/01/2010) to 2455927.5 (01/01/2012) in Julian days.

Similarly, a single query can introduce multiple values for a given parameter. ADQL provides standard operations on parameters to combine possible conditions (and, or, like…) as well as parentheses. Standard ADQL wildcards are also implemented.

*target_name like 'Mars' or target_name like 'Venus'*

Return data on either Mars or Venus.

The current ADQL standard is however causing troubles here: the query on 1P above will also provide results on 11P, 21P, etc, which are different comets.

Another limitation of ADQL forces to provide and query most parameters in lower-case, which leads to inconsistencies as detailed in section 3.3. Case sensitive parameters are: target names, URLs, filenames, "species" and all non-standard parameters (i.e., defined for a particular service and not listed here). Those are currently handled via the ivo_nocasematch function in DaCHS, and there are plans to implement a similar system in VO-Dance.

### 5.2 Service response

The response of the service is formatted as a VOTable, which must comply with the VOTable standard, version 1.2 or higher.

Following the TAP protocol, the response contains information about the service, the query, and the epn_core view; it also contains the data itself or links to data files.

The VOTable must contain a RESOURCE element with the attribute *type="results"* containing a single TABLE element with the results of the query. Additional RESOURCE elements may be present, but the usage of any such element is not defined here and the TAP client may not use them.

The Resource element includes INFO elements providing: the URL of the data server, the EPN-TAP query and its status, descriptions of the service and table, and a credit note. The content of the INFO elements is a message suitable for display to the user.

The Resource element also includes a TABLE element providing a description of the epn_core view columns, with the fields name, data type, unit, and UCD. This is followed by a data area containing the subset of rows from the epn_core view that match the query. All parameters in the view are therefore available to the client. The data itself is either linked with an access URL or directly embedded in the response VOTable, depending on the service view. The issue of possible format conversion is left to the client or visualization tools. If no result fulfills the query, the TABLE element must be present and empty (i.e., the TABLE element has no DATA element). Otherwise, it may be encoded in binary using the base64 scheme.

# Conclusion

The EPN-TAP protocol provides a consistent way to query many services of interest for Planetary Science in the fields of observations, simulations, and laboratory measurements. Although similar to ObsCore in many respects, EPN-TAP has broader focus but is not intended to replace ObsCore – rather to complement it to distribute Planetary Science content.

At the moment of writing, the protocol is still in test phase but very close to completion. It is already discussed in IVOA and IPDA working groups, and will be the default protocol implemented on coming Europlanet services.

Future steps of development will include:
- The improvement of reference lists for the string parameters. In some cases, these lists will be elaborated from existing but incomplete or contradicting references. Coordinate systems in use in the Solar System and instrument hosts appear to be the most sensitive.
- Specific UCDs are required to describe the quantities routinely measured in this field, in

particular concerning measurements in reflected light and particle properties. This is currently discussed in the IVOA working groups.
- An evolution of ADQL to overcome present difficulties related to case handling and multiple valued fields.

# Acknowledgements

This work has been conducted in the frame of Europlanet-RI JRA4 work package.
The EuroPlaNet-RI project is funded by the European Commission under the 7th Framework Program, grant #228319 "Capacities Specific Programme". Additional funding was provided in France by the Association Spécifique Observatoire Virtuel / INSU.

Table 1: EPNCore parameters

| Name | Class | Unit | Description | UCD |
|---|---|---|---|---|
| **Mandatory parameters** | | | | Must be present |
| index | Long | | Internal table row index | meta.id |
| resource_type | String | | Can be dataset or granule | meta.id;class |
| dataset_id | String | | Dataset identification & granule reference | meta.id;meta.dataset |
| dataproduct_type | String | | Organization of the data product, from enumerated list | meta.id;class |
| target_name | String | | Standard name of target (from a list depending on target type), case sensitive | meta.id;src |
| target_class | String | | Type of target, from enumerated list | src.class |
| time_min | Float/double | d | Acquisition start time (in JD) | time.start |
| time_max | Float/double | d | Acquisition stop time (in JD) | time.end |
| time_sampling_step_min | Float | s | Min time sampling step | time.interval;stat.min |
| time_sampling_step_max | Float | s | Max time sampling step | time.interval;stat.max |
| time_exp_min | Float | s | Min integration time | time.duration;stat.min |
| time_exp_max | Float | s | Max integration time | time.duration;stat.max |
| spectral_range_min | Float | Hz | Min spectral range (frequency) | em.freq;stat.min |
| spectral_range_max | Float | Hz | Max spectral range (frequency) | em.freq;stat.max |
| spectral_sampling_step_min | Float | Hz | min spectral sampling step | em.freq.step;stat.min (not standard) |
| spectral_sampling_step_max | Float | Hz | Max spectral sampling step | em.freq.step;stat.max (not standard) |
| spectral_resolution_min | Float | Hz | Min spectral resolution | spect.resolution;stat.min |
| spectral_resolution_max | Float | Hz | Max spectral resolution | spect.resolution;stat.max |
| c1min | Float | deg | Min of first coordinate | Pos;stat.min |
| c1max | Float | deg | Max of first coordinate | Pos;stat.max |
| c2min | Float | deg | Min of second coordinate | Pos;stat.min |
| c2max | Float | deg | Max of second coordinate | Pos;stat.max |
| c3min | Float | | Min of third coordinate | Pos;stat.min |
| c3max | Float | | Max of third coordinate | Pos;stat.max |
| c1_resol_min | Float | deg | Min resolution in first coordinate | Pos.resolution;stat.min (not standard) |
| c1_resol_max | Float | deg | Max resolution in first coordinate | pos.resolution;stat.max (not standard) |
| c2_resol_min | Float | deg | Min resolution in second coordinate | pos.resolution;stat.min (not standard) |
| c2_resol_max | Float | deg | Max resolution in second coordinate | pos.resolution;stat.max (not standard) |
| c3_resol_min | Float | | Min resolution in third coordinate | pos.resolution;stat.min (not standard) |
| c3_resol_max | Float | | Max resolution in third coordinate | pos.resolution;stat.max (not standard) |
| spatial_frame_type | String | | Flavor of coordinate system, defines the nature of coordinates | pos.frame |

| | | | | | |
|---|---|---|---|---|---|
| incidence_min | float | | Min incidence angle (solar zenithal angle) | pos.incidenceAng;stat.min (not standard) | |
| incidence_max | float | | Max incidence angle (solar zenithal angle) | pos.incidenceAng;stat.max (not standard) | |
| emergence_min | float | | Min emergence angle | pos.emergenceAng;stat.min (not standard) | |
| emergence_max | float | | Max emergence angle | pos.emergenceAng;stat.max (not standard) | |
| phase_min | float | | Min phase angle | pos.phaseAng;stat.min (not standard) | |
| phase_max | String | | Max incidence angle | pos.phaseAng;stat.max (not standard) | |
| instrument_host_name | String | | Standard name of the observatory or spacecraft | meta.class | |
| instrument_name | String | | Standard name of instrument | meta.id;instr | |
| measurement_type | String | | UCD(s) defining the data | meta.ucd | |
| **Optional parameters** | | | | Must be used in this sense if present | |
| access_url | String | | URL of the data file, case sensitive | meta.ref.url | |
| access_format | String | | File format type | meta.code.mime | |
| access_estsize | Integer | kB | Estimate file size in kB | phys.size;meta.file | |
| preview_url | Integer | | URL of a preview image | meta.id;meta.file | |
| native_access_url | String | | URL of the data file in native form, case sensitive | meta.ref.url | |
| native_access_format | String | | File format type in native form | meta.code.mime | |
| file_name | String | | Name of the data file only, case sensitive | meta.ref.url | |
| species | String | | Identifies a chemical species, case sensitive | phys.composition.species (not standard) | |
| element_name | String | | Secondary name (can be standard name of region of interest) | meta.id | |
| Reference | String | | Bibcode or other bilbio id | meta.bib | |
| ra | Float | | Right ascension | pos.eq.ra;meta.main | |
| dec | Float | | Declination | pos.eq.dec;meta.main | |
| ls | Float | | Solar longitude | | |
| target_distance | Float | km | Observer-target distance | pos.distance | |
| particle_spectral_type | String | | | | |
| particle_spectral_range_min | Float | | | | |
| particle_spectral_range_max | Float | | | | |
| particle_spectral_sampling_step_min | Float | | | | |
| particle_spectral_sampling_step_max | Float | | | | |
| particle_spectral_resolution_min | Float | | | | |
| particle_spectral_resolution_max | Float | | | | |
| **Relative to service / Table header** | | | | Can be used as optional parameters | |
| processing_level | Integer | | CODMAC calibration level | meta.code;obs.calib | |
| publisher | String | | Resource publisher | meta.name | |
| reference | String | | Reference publication | meta.bib | |
| service_title | String | | Title of resource | meta.id | |
| spatial_coordinate_description | String | | Indicates exact spatial frame | | |
| spatial_origin | String | | Defines the frame origin | | |

| time_origin | String | | Defines where the time is measured | |
| target_region | String | | Type of region of interest | meta.id;class |

Table 2: Spatial Frame Types

| celestial | 2D angles on the sky: Right Ascension c1 and Declination c2 + possibly distance from origin c3. Although this is a special case of spherical frame, the order is different. |
|---|---|
| body | 2D angles on a rotating body longitude: c1 and latitude c2 + possibly altitude/depth c3. Default is IAU 2009 planetocentric convention, east-handed [12]. |
| cartesian | (x,y,z) as (c1,c2,c3). This includes spatial coordinates given in pixels. |
| cylindrical | (r, theta, z) as (c1,c2,c3). Angles are defined in degrees. |
| spherical | (r, theta, phi) as (c1,c2,c3). Angles are defined as in usual spherical systems (E longitude, zenithal angle/colatitude), in degrees. If the data are related to the sky, "celestial" coordinates with RA/Dec must be used. |
| healpix | (H, K) as (c1,c2) |

Table 3: Target Types

| From IAU list [18] | asteroid, dwarf_planet, planet, satellite |
|---|---|
| Extra EPN-TAP types | comet, exoplanet, interplanetary_medium, ring, sample, sky, spacecraft, spacejunk, star |

Table 4: Data Product Types

| EPN-TAP value | Type | Description |
|---|---|---|
| im | image | Scalar field with two spatial axes, or association of several such fields. Maps of planetary surfaces are considered as images. |
| sp | spectrum | Measurements organized primarily along a spectral axis, e.g., a series of radiance spectra. |
| ds | dynamic_spectrum | Consecutive spectral measurements through time, organized as a time series. |
| sc | spectral_cube | Sets of spectral measurements with 1 or 2D spatial coverage, e.g., imaging spectroscopy. |
| pr | profile | Scalar or vectorial measurements along 1 spatial dimension, e.g., atmospheric profiles, atmospheric paths, sub-surface profiles, etc |
| vo | volume | Other measurements with 3 spatial dimensions, e.g., internal or atmospheric structures. |
| mo | movie | Sets of chronological 2D spatial measurements |
| cu | cube | Multidimensional data with 3 or more axes, e.g., all that is not described by other 3D data types such as spectral cube or volume. |
| ts | time_series | Measurements organized primarily as a function of time (with exception of dynamical spectra and movies, i.e. usually a scalar quantity). |
| ca | catalog | Lists of events, catalogs of object parameters, lists of features… The primary key may be a qualitative parameter (name, ID, etc). |
| sv | spatial_vector | List of summit coordinates defining a vector, e.g., vector information from a GIS, spatial footprints, etc |

Table 5: Processing levels

| CODMAC level / EPN-TAP value | PSA level | NASA level | PRODUCT_TYPE (PDS/PSA) | ObsTAP | Description |
|---|---|---|---|---|---|
| 1 (raw) | 1a | | UDR | Level 0 | Unprocessed Data Record (low-level encoding, e.g. telemetry from a spacecraft instrument. Normally available only to the original team) |
| 2 (edited) | 1b | 0 | EDR | Level 1 | Experiment Data Record (often referred to as "raw data": decommutated, but still affected by instrumental effects) |
| 3 (calibrated) | 2 | 1A | RDR | Level2 | Reduced Data Record ("calibrated" in physical units) |
| 4 (resampled) | | 1B | REFDR | | Reformatted Data Record (mosaics or composite of several observing sessions, involving some level of data fusion) |
| 5 (derived) | 3 | 2-5 | DDR | Level3 | Derived Data Record (result of data analysis, directly usable by other communities with no further processing) |
| 6 (ancillary) | | | ANCDR | | Ancillary Data Record (extra data specifically supporting a data set, such as coordinates, geometry, etc) |